\documentclass[aps,prl,preprint]{revtex4}
\usepackage{graphicx}
\usepackage{color}
\begin{document}

\title{
Finite electric field effects in the large perpendicular magnetic anisotropy 
surface Pt/Fe/Pt(001)}

\author{Masahito Tsujikawa$^1$ and Tatsuki Oda$^2$}
\affiliation{
$^{1}$Graduate School of Natural Science and Technology,
Kanazawa University, Kanazawa 920-1192, Japan\\
$^{2}$Institute of Science and Engineering,
Kanazawa University, Kanazawa 920-1192, Japan}

\date{\today}

\begin{abstract}
We have investigated crystalline magnetic anisotropy in the electric field 
(EF) for the Fe-Pt surface which have a large perpendicular anisotropy, 
by means of the first-principles approach. The anisotropy 
is reduced linearly with respect to the inward EF, 
associated with the induced spin density around the Fe layer. 
Although the magnetic anisotropy energy (MAE) density reveals 
the large variation around the atoms, the intrinsic contribution to 
the MAE is found to mainly come from the Fe layer. 
\end{abstract}


\maketitle



Magnetoelectric properties in the solid state are attractive for 
the spintronics applications.
Through the spin-orbit interaction (SOI) the magnetic and electric 
properties are connected and consequently, the electric field (EF) 
allows us to manipulate the magnetic properties \cite{ChibaNature08} 
and the magnetic field could control the electric polarization of 
materials \cite{Kimura03}. 
Such properties may be allowed in the system which looses 
the time-reversal and space-inversion symmetries, e.g., 
surface magnetization systems.
One of the important problems is to control the magnetic anisotropy 
with the voltage or the EF. 
The EF induced variation of coercivity has been observed in the 
ferromagnetic semiconductor \cite{ChibaScience03} 
and the large perpendicular magnetization metallic surface 
(FePt and FePd thin layers) \cite{Weisheit07}.
The strength of magnetic anisotropy needs in the nanoscale 
device so as not to loose the magnetic memory by the thermal fluctuation 
and, meanwhile, makes difficulties in reversing the magnetization. 
The large experimental observation of EF effect on coercivity has been 
reported \cite{Sahoo07}.
The direct estimation of magnetic anisotropy energy (MAE) was performed 
for the thin film, Au/Fe/MgO \cite{Maruyama09}. 
Recent theoretical works on magnetoelectric properties in the thin 
iron films have explained the variation of MAE \cite{Duan08APL,Duan08PRL}, 
based only on a simplified Bruno's relation (relationship between the 
MAE and the atomic orbital magnetic moments) \cite{Bruno89}.

In order to obtain a built-up technology, the stable theoretical 
background for the EF effect is required in the basis of realistic 
electronic structures. 
The present theoretical work shows that the surface accumulated charge 
modifies the strength of magnetic anisotropy through the modulation of 
electronic state at the magnetic layer. 
We will discuss origins of the change and intrinsic contributions 
to the MAE. 


We have carried out first-principles electronic structure calculations 
\cite{Oda98} which employ ultrasoft pseudopotentials \cite{Laasonen93} 
and planewave basis.
Except for imposing the EF, the details about the method and the models 
are the same as in the previous study for Pt/Fe/Pt(001) 
in no EF \cite{TsujikawaPRB08}.
These systems have four atomic Pt layers for the substrate 
and the atomic positions of the three bottom layers were 
fixed to the appropriate values of the bulk fcc Pt.
The other atoms were relaxed using the calculated atomic forces under 
the zero EF. All the atomic positions were fixed for finite EFs in MAE 
calculations. 
The atomic displacements induced by the EF may be a future problem 
for the magnetoelectric physics and the MAE estimation. 
In order to impose the EF, we have applied the scheme of effective screening 
medium (ESM) developed by Otani and Sugino \cite{Otani06}. 
In the present work, the ideal conductor was placed away from the Fe layer 
by 7.0 \AA. Some tiny number of electrons was added in the slab for induction 
of the EF and at the same time induction of the counterpart charge at 
the ESM surface. The strength of EF was estimated at the front of ESM.

The MAE was estimated from the total energy of the system 
with the in-plane magnetization ([100] direction) with respect to that 
of the out-of-plane (perpendicular) magnetization ([001] direction). 
In the present work the MAE density, $D({\bf r})$, has been introduced 
to understand the local contribution of MAE from a coarse-grain region 
of real space. 
The total energy $E_{\rm tot}^{{\bf m}}$, where ${\bf m}$ specifies the 
magnetization direction of system, may formally be divided to the local 
point in real space; $E_{{\bf m}}({\bf r})$, where 
$\int_{\rm cell} E_{{\bf m}}({\bf r}) d{\bf r} = E_{\rm tot}^{{\bf m}}$. 
The atomic contribution 
in the total energy (non-local part from the pseudopotential, etc.) was 
redefined as a Gaussian form function centered at the atomic position. 
The MAE density was defined by 
$D({\bf r})=E_{[100]}({\bf r})- E_{[001]}({\bf r})$ and the MAE is 
alternatively obtained by $\int_{\rm cell} D({\bf r}) d{\bf r}$. 
The atomic contributions of MAE were estimated by integrating 
$D({\bf r})$ within the atomic sphere with the radius of 1.3 \AA \ and 
the layer contributions within the layer with 
the plane boundaries determined with the midpoint of atomic coordinates along 
the $c$-direction or the surface normal. The $z$-dependent MAE density 
$\Delta(z)$ was obtained by summing $D({\bf r})$ up within the in-plane unit 
cell in the fixed plane normal to the surface. 
Further, we introduced another integrated MAE density; 
$T(z)=\int_{-\infty}^{z} \Delta(z^{\prime}) dz^{\prime}$. 
This function helps us to capture the feature of $\Delta(z)$.
The functions introduced above were calculated at a given EF (${\cal E}$);
$D({\bf r},{\cal E})$, $\Delta(z,{\cal E})$, and $T(z,{\cal E})$.


The bulk L1$_{0}$-FePt has a magnetic anisotropy along  
the $c$-direction. The total MAE is divided to contribution of each 
atom (1.96 meV and $-$0.01 meV in Fe and Pt atoms, respectively) and 
the interstitial region (0.67 meV/f.u.) \cite{Oda05,Ravindran01}.
This total MAE is also partitioned to the Fe and Pt layers along the 
$c$-direction, resulting in 2.83 meV and $-$0.21 meV in Fe and Pt 
layers, respectively. 
These results, though the values associated with Pt are 
small and negative, have never expressed a minor contribution to 
magnetic anisotropy from Pt atoms. 
The intra-atomic spacial variation around Pt atoms is remarkable,
reflected from a large spin-orbit interaction on Pt atoms and 
the hybridization of Pt $5d$ with Fe $3d$ are responsible to the 
large magnetic anisotropy of such systems \cite{Komelj06}. 
The spacial partition on MAE becomes very interesting in the surface
systems and also under the EF, because the potential gradient against
electrons is changed and the electronic structure is modulated, thus,
which will result in the modulation of effective spin-orbit parameter
 \cite{Oguchi09}.

\begin{figure}[tb]
  \begin{center}
     \vspace*{1cm}
     \begin{tabular}{cc} 
       \rotatebox{00}{\resizebox{80mm}{!}
           {\includegraphics{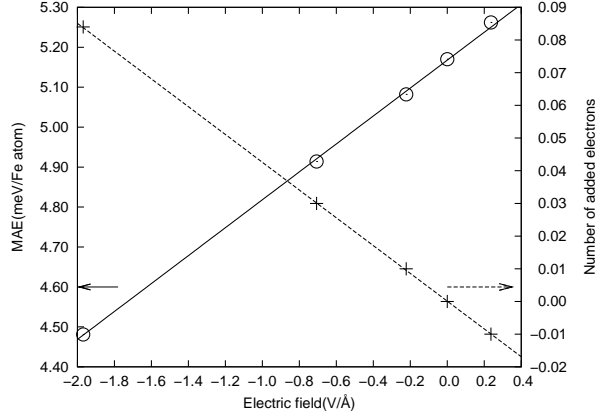}}} 
     \end{tabular}
     \caption{Magnetic anisotropy energies (MAEs) (circles) 
            and the number of added electrons (crosses) as a function of 
            the electric field for Pt/Fe/Pt(001).
            The lines are obtained by the least squares fit 
            to the set of data.
            }
        \label{figMAE}
  \end{center}
\end{figure}

\begin{figure*}[htb]
  \begin{center}
     \begin{tabular}{ccc}
       \rotatebox{00}{\resizebox{80mm}{!}
           {\includegraphics{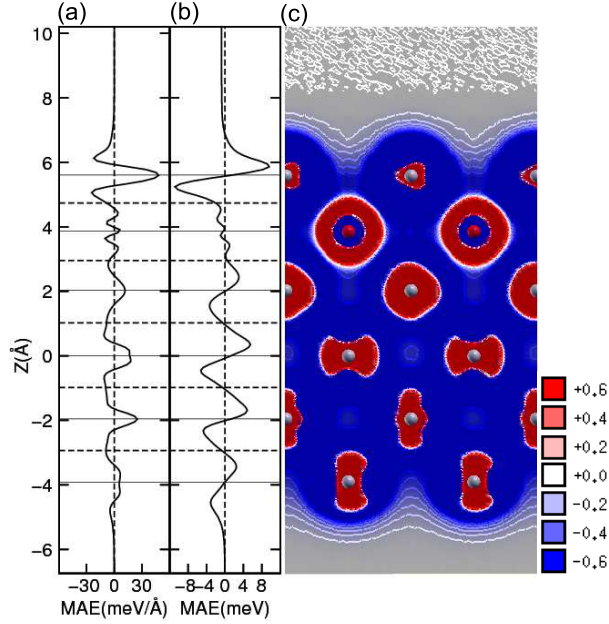}}}
       &\ \ \ \ &
     \end{tabular}
       \caption{(color online). Electric field change of the magnetic 
                anisotropy energy (MAE) densities along $z$-coordinate 
                and the contour (meV/${\rm \AA}^3$) 
                maps, (a)-(c) for Pt/Fe/Pt(001).
                The horizontal solid and dashed lines in (a)(b) indicate 
                the atomic positions and the layer's boundaries, respectively,
                for better visualization.
                The map is shown in the (110) plane 
                on the Fe (red ball) and Pt (gray ball) atoms.
                }
     \label{figMAEmap}
  \end{center}
\end{figure*}

\begin{figure}[htb]
  \begin{center}
       \rotatebox{00}{\resizebox{85mm}{!}
           {\includegraphics{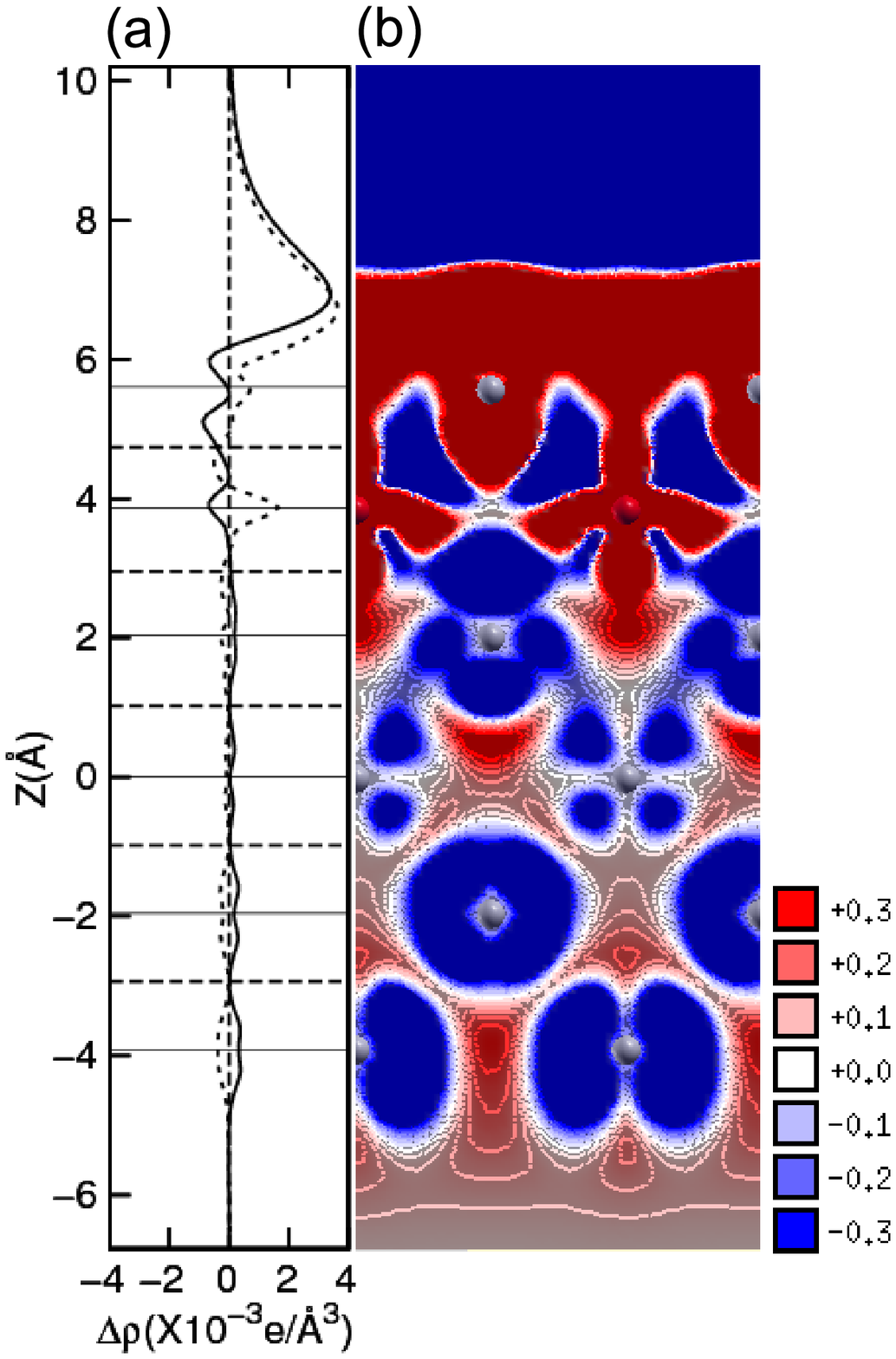}}}
       \caption{(color online). The induced majority- (solid curves) and 
                minority- 
                (dashed curves) spin charge densities along $z$-coordinate 
                and the contour maps of induced spin densities
                ($\times 10^{-3} {\rm e/\AA}^3$) (a)(b) for Pt/Fe/Pt(001).
                The magnetization direction used for the figure drawing 
                is perpendicular to the surface plane.
                See the caption at Fig. \ref{figMAEmap} for other 
                explanations.
                }
       \label{figdensity}
  \end{center}
\end{figure}

In Pt/Fe/Pt(001), the MAE contributes mainly from Fe layer 
at zero field (5.50 meV/Fe), amounting to 121 \% of the 
total (5.21 meV/Fe).  The main negative contribution  comes from the 
capping Pt layer ($-$0.85 meV/Fe). Imposing the inward EF the MAE 
decreases, associated with increase of the number of added electrons. 
The similar decrease 
has been reported in the literature \cite{Weisheit07}, 
where the decrease of MAE speculatively coincides with the bandfilling 
in the L1$_{0}$-FePt \cite{Daalderop91}.
Figure \ref{figMAE} presents the EF dependence of MAEs in Pt/Fe/Pt(001), 
associated with the number of added electrons in the unit cell.
The method for imposing EFs \cite{Otani06} almost unchanged 
($\sim 0.01 {\rm meV/Fe}$) the MAE at zero field, compared with the previous 
estimation with the simple periodic slab approach \cite{TsujikawaPRB08}. 
The dependence of the EF on the number of added electrons is almost linear. 
This slope rate corresponds to the dielectric constant for vacuum 
($\varepsilon_{0}$); $\sim 1.00 \varepsilon_{0}$. 
The decrease rate of MAE with respect to the inward EF is estimated to be 
0.35 meV per Fe atom per V/\AA, corresponding to the surface MAE of 72 fJ/Vm. 
This rate is larger than the previous theoretical estimation for 
the metallic surface of iron by the factor of 3.8 ($\sim$ 19 fJ/Vm) 
\cite{Duan08PRL}.
Moreover, although the substrate of system is different, this slope 
could explain a partial contribution of the change of MAE by voltage 
in the experimental measurement \cite{Maruyama09}.

In Figs. \ref{figMAEmap}(a)-(c) the EF-induced MAE densities,
$\Delta(z,{\cal E}_{1})\!-\!\Delta(z,0)$,
$T(z,{\cal E}_{1})\!-\!T(z,0)$, and
$D({\bf r},{\cal E}_{1})\!-\!D({\bf r},0)$,  
where ${\cal E}_{1}\!\!=\!\! 
-1.97 {\rm V}/{\rm \AA}$, are shown along 
$z$-direction in Pt/Fe/Pt(001). In these figures, the MAE density has 
large strengths around Pt atoms, which reflects the large SOI at them.
However, such contributions to the MAE seem to cancel out around Pt atoms 
and, as shown in Fig. \ref{figMAEmap}(b), the intrinsic contribution to 
the total induced variation on MAE is raised around the Fe layer.
This indicates that the local electronic structure centered at the Fe 
layer is intrinsically important for the change of MAE.

To see the relationship between the variation of MAE and the EF 
in more details, we have calculated the induced spin density, 
the energy-dependent density of states for five $3d$ angular-dependent 
orbitals 
on Fe atom, and the orbital-specified band dispersions \cite{TsujikawaPRB08}. 
Figures \ref{figdensity} (a) and (b) show the induced majority- and 
minority-spin densities along $z$-direction and the induced spin density 
map, respectively, at ${\cal E}= -1.97 {\rm V/\AA}$ in Pt/Fe/Pt(001). 
At Fe the minority-spin density is induced and this induction is found 
to result in a partial cancellation of the increase of 
$d_{3z^2-r^2}$ and $d_{x^2-y^2}$ components against the decrease of 
$d_{xz}$ and $d_{yz}$. 
The latter is associated with the charge accumulation at outside of the 
capping layer through the orbital hybridization between Fe $3d$ and Pt $5d$, 
which implies a subtle bonding reduction. As a result, the $d_{xz}$ and 
$d_{yz}$ components grow just above the Fermi level. 
These EF modulations in electronic structure around the Fermi level
can be related with the decrease of MAE by using the second order 
perturbative consideration \cite{Wang93model};
\begin{equation}
{\rm MAE}   \ \ \propto \ \ 
\sum_{\bf k}\sum_{o,u}
\frac{\left|\langle {\bf k}o | \ell_z | {\bf k}u \rangle \right|^2
               -\left|\langle {\bf k}o | \ell_x | {\bf k}u \rangle \right|^2}
               {\varepsilon_{{\bf k}u}-\varepsilon_{{\bf k}o}},
\label{eqsecond}
\end{equation}
where ${\bf k}o$ and ${\bf k}u$ specify the occupied and unoccupied states 
with the wave vector {\bf k} and $\ell_{\alpha}(\alpha=x,z)$ the angular 
momentum operators.
The EF modulation mentioned above increases the couplings between occupied 
and unoccupied states (couplings of $\langle 3z^{2}\! -\! r^{2}||yz \rangle$ 
and $\langle yz||x^{2}\!-\!y^{2} \rangle$) through the $\ell_{x}$ 
operator and, thus, reduces the MAE \cite{Wang93model}.

In Table \ref{tablemae}, the spin and orbital moments are reported. 
These quantities change linearly with the EF in Pt/Fe/Pt(001).
One would explain the MAE from the orbital magnetic moments, while  
it is not probable that the simplified Bruno's relation applies to the 
FePt alloy (see Eq.(9) in \cite{Ravindran01}) without the spin flip 
contribution \cite{Bruno89}. 
This is because the Pt does not have any large exchange splitting.
As implied by the difference of atomic orbital magnetic moments on 
Fe and Pt atoms in Table \ref{tablemae}, the MAE may be supposed to 
mainly come from the Pt atoms. This picture is in contradiction with 
the feature obtained from the MAE density (Fig. \ref{figMAEmap}). 
In this context, to give a reasonable explanation in the relation with 
the orbital magnetic moment and the MAE, the application of 
a general Bruno's relation is required, accompanied with the spin flip 
contribution and the interstitial (inter-atomic) contribution to the MAE.

\begin{table}[b]
\begin{center}
\caption{Magnetoelectric properties in Pt/Fe/Pt(001);
magnetic anisotropy energies(MAEs) at zero field and their gradients with 
respect to the electric field(${\cal E}$), the spin magnetic moment (SMM) 
from the three surface layers in the [001] magnetization, 
and the difference of atomic orbitals magnetic moments(DOMM) between 
the [001] and [100] magnetizations, $M_{\rm orb}[001]-M_{\rm orb}[100]$.}
\label{tablemae}
\begin{tabular}{lc}
\hline
                           & Pt/Fe/Pt(001)        \\
\hline
MAE at ${\cal E}=0$        &  5.21\ (11)$^{a}$    \\
MAE slope rate             &  0.35\ (72)$^{b}$    \\
SMM  at ${\cal E}=0$       &  3.72      $^{c}$    \\
SMM slope rate             &  0.0065(0.99) $^{d}$ \\
Fe DOMM at ${\cal E}=0$    &  0.0080    $^{c}$     \\
Fe DOMM slope rate         &  $-$0.0037 ($-$0.56)  $^{d}$   \\
Pt(c) DOMM at ${\cal E}=0$ &  $-$0.0417   $^{c}$            \\
Pt(c) DOMM slope rate      &  $-$0.0024 ($-$0.36)  $^{d}$   \\
\hline
\end{tabular}
\\
$^{a}$ in meV/Fe (J/m$^{2}$),\ $^{b}$ in meV/Fe per V/\AA\ (fJ/Vm),\\ 
$^{c}$ in $\mu_{\rm B}$,\  $^{d}$ 
in $\mu_{\rm B}$ per V/\AA (10$^{-18}$ Gm$^{2}$/V)
\end{center}
\end{table}

Imposing the EF, the number of electrons changes in a few layers
of the surface due to a screening effect of metal,
as shown Fig. 3(a). This feature should be realized
also in the experiment \cite{Weisheit07}. If the change of coercivity
is assumed to be attributed to the one or two magnetic layers of the 
metal surface in the experiment, the MAE of these layers could reduce
by 10 $\sim$ 30 \% for the EF of $-0.03$V/\AA.
This reference to the experiment provides a good agreement with
our decrease of MAE (13 \% in Pt/Fe/Pt(001)) by the EF of $-0.030$V/\AA \ 
which is scaled by the dielectric constant of the experimental substrate
on metal surface ($-1.97$V/\AA \ on vacuum). 


In summary, we have studied the EF dependence of MAE for the large 
perpendicular magnetic anisotropy surface system. 
The MAE linearly decreases with the inward EF in Pt/Fe/Pt(001). 
By analysis of the induced MAE density, the large variation around Pt 
atoms was revealed and the intrinsic contribution to the MAE was verified 
to mainly come from the Fe layer.
The analysis of MAE density provides a promising tool for 
pursuing the spacial contribution of nanoscale structures, 
regardless of the knowledge of orbital magnetic moments. 
The present study indicates that the relative modification in the 
electron filling of each $3d$ orbital by the EF, resulting in the 
accumulated charge at the magnetic layer, causes the variation of MAE.


\end{document}